\def\cerntp{1}

\documentstyle[12pt,epsfig,ifthen,citesort]{article}

\newcommand{\vdate}{August 1996}
\newcommand{\cernnr}{96--236}

\newlength{\dinwidth}                       
\newlength{\dinmargin}                      
\ifnum\cerntp=0
\fi
\newcommand{\gsim}{\raisebox{-0.07cm}{\, $\stackrel{>}{{\scriptstyle
 \sim}}$\, }}
\newcommand{\lsim}{\raisebox{-0.07cm}{\, $\stackrel{<}{{\scriptstyle
 \sim}}$\, }}
\newcommand{\titletextCERN}
{
The Extraction of the Gluon Density\\ 
from Jet Production
in Deeply Inelastic Scattering
}

\newcommand{\titletextPL}
{
The Extraction of the Gluon Density \\ 
from Jet Production
in Deeply Inelastic Scattering\\
}

\newcommand{\abstracttext}
{
The prospects of a direct extraction of the proton's gluon density in 
next-to-leading order via jet rates in deeply inelastic scattering
are studied.
The employed method is based on the Mellin transform, and can be 
applied, in principle, to all infra-red-safe observables of hadronic 
final states.
We investigate the dependence of the error band on the extracted gluon 
distribution on the statistical and systematic error of the data.
}

\begin{document}


\ifnum\cerntp=1
   
   \thispagestyle{empty}
   
   \renewcommand{\thefootnote}{\fnsymbol{footnote}}
   \setcounter{footnote}{0}
   
   \begin{flushright}
   {
   \unitlength 1mm
   \begin{picture}(10,10)
   \put(0,0){CERN--TH/\cernnr}
   \put(0,-5){WUE-ITP-96-015}
   \put(0,-10){hep-ph/9609446}
   \end{picture}
   \rule{2cm}{0mm}
   }
   \end{flushright}
   \vspace{1.5cm}
   
   \begin{center}
   
   {\Large\bf \titletextCERN$\!\!$\footnote[3]{\it Contribution to the
   workshop ``Future Physics at HERA'' at DESY, Hamburg, 1995/96; to be
   published in the proceedings.}\\}
   
   
   \vspace{1.0cm}

\footnote[0]{\it Electronic mail addresses:
Dirk.Graudenz@cern.ch, Hampel@desy.de,
avogt@physik.uni-wuerzburg.de}

{\bf 
D.~Graudenz$^1$\footnote{\it WWW URL:
http://wwwcn.cern.ch/$\sim$graudenz/index.html}, 
M.~Hampel$^2$, 
A.~Vogt$^3$}

\vspace{2mm}  

\vspace{3mm}
{{}$^1 \it $Theoretical Physics Division, CERN\\ CH--1211 Geneva 23, 
 Switzerland}

\vspace{3mm}
{{}$^2 \it $I.\ Physikalisches Institut, RWTH Aachen\\
D--52056 Aachen, Germany}

\vspace{3mm}
{{}$^3 \it $Institut f\"ur Theoretische Physik,
 Universit\"at W\"urzburg\\ Am Hubland, D--97074 W\"urzburg, Germany}

   \end{center}  
   
   \vspace{1.0cm}

   \begin{center}
   {\bf Abstract}
   \end{center}
   
   \hspace{4mm}
   \abstracttext 
   
   \vfill
   \noindent
   CERN--TH/\cernnr\\
   \vdate
   
   \clearpage
   \setcounter{page}{1}
   
\fi
   

\vspace*{1cm}
\begin{center}  \begin{Large} \begin{bf}
\titletextPL
  \end{bf}  \end{Large}
  \vspace*{5mm}
  \begin{large}
D.~Graudenz$^a$, M.~Hampel$^b$, A.~Vogt$^c$\\
  \end{large}
\end{center}
$^a$ Theoretical Physics Division, CERN, CH--1211 Geneva 23, Switzerland\\
$^b$ I.~Physikalisches Institut, RWTH Aachen,
     D--52056 Aachen, Germany\\
$^c$ Institut f\"ur Theoretische Physik, Universit\"at W\"urzburg, 
     D--97074 W\"urzburg, Germany\\
%


\begin{quotation}
\noindent
{\bf Abstract:}
\abstracttext
\end{quotation}

\renewcommand{\thefootnote}{\arabic{footnote}}
\setcounter{footnote}{0}

\section{Introduction}

The extraction of the proton's gluon density over a wide kinematical 
range is one of the central issues at HERA. This distribution is 
important for phenomenological applications at $pp$-colliders, as well as at 
small~$x$, $x \lsim 0.005$ say, for studies related to parton dynamics, 
since higher-order terms of the type $[\alpha_s \ln (1/x)]^k$ in the 
perturbative expansion may become important here. At HERA, this region 
is mainly covered by the scaling violations and (with increasing 
luminosity) more directly by the charm content of the structure 
function $F_2$.
The classical gluon constraint at larger $x$ ($x>0.01$),
direct photon production in $pp$ collisions, is plagued by sizeable 
theoretical uncertainties, see Ref.~\cite{1}. It is therefore desirable 
to have a direct determination of the gluon density from HERA also at 
$x \gsim 0.01$, complementing the observables so far employed in global 
fits (see, for example, Ref.~\cite{2}).

Experimental analyses of hadronic final states use very complicated
sets of cuts. The theoretical predictions for the employed
infra-red-safe observables are in general obtained by means of 
time-consuming Monte-Carlo integrations of parton cross-sections.
This is contrary to the case of DIS structure functions, for example, where
the convolution kernels can be given in a compact analytical form.
Parametrizations of the gluon density involve some free parameters 
--- for a decent description at, say, a scale of a few GeV$^2$ at least 
$3 - 4$ parameters are necessary \cite{1,2} --- so that a fitting 
procedure with a large number of iterations is unavoidable. 
This leads to the problem of the repeated evaluation of the theoretical 
cross-section for various gluon density parametrizations. It turns out 
that the direct method is not easily feasible due to its prohibitive
need of computer time. 
A method based on the Mellin transform technique\footnote{
The Mellin transform
$F_n=\int_0^1\mbox{d}x\, x^{n-1}\,F(x)$ maps a convolution
$\sigma(x_B)=\int_{x_B}^1 (\mbox{d}\xi/\xi) f(\xi)\,\sigma^p(x_B/\xi)$ of a 
parton density $f$ and a parton-level cross-section $\sigma^p$ into
the product $\sigma_n=f_n\,\sigma^p_n$ of the respective moments.
For a specific set of acceptance cuts, the time-consuming Monte Carlo
calculation of the moments $\sigma^p_n$
has to be done only once. The 
cross-section~$\sigma$ can be evaluated repeatedly for varying 
parton densities by means of an inverse Mellin transform.
A restriction of the method is that the factorization scale 
in every bin of analyzed data is assumed to be constant.
However, owing to scale-compensating terms in the parton-level cross-section,
the procedure is always accurate to the order of perturbation theory
for which the cross section is calculated.
For details, in particular on how to treat the non-factorizable case,
see Ref.~\cite{3}.
}
to circumvent this problem has been presented in Ref.~\cite{3}.
It allows the rapidly repeated evaluation of the convolution of a 
cross-section and a parton density, even if the cross-section is not of the
Mellin-factorizing form. In fact, this is always the case when 
acceptance cuts are applied, even for ``factorizing'' jet definition
schemes \cite{4}.

\begin{figure}[tbp] \unitlength 1mm
\begin{center}
\begin{picture}(160,45)
\put(-0.3,0){\epsfig{file=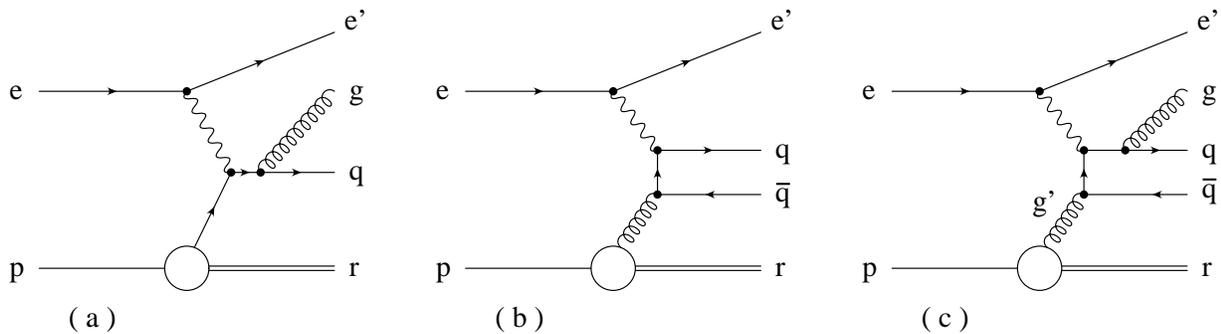,width=16cm}}
\end{picture}
\end{center}
\vspace{-5mm}
\caption[]
{{\it Generic Feynman diagrams for the leading-order processes of
QCD Compton scattering (a) and photon--gluon fusion (b), and an 
example for a diagram corresponding to a next-to-leading order real 
correction (c).}}
\label{figref1}
\end{figure}

Jet observables are particularly suitable for the extraction of the 
gluon density. The reason for this is that the lowest-order 
gluon-induced subprocess is the photon--gluon fusion process 
(Fig.~\ref{figref1}b), which can give rise to (2+1) jet final states. 
The cross-section of the competing quark-initiated 
``QCD Compton''-subprocess (Fig.~\ref{figref1}a) is either small (at 
small~$x$) or well-known (at large~$x$). Hence it can be determined
theoretically and then subtracted from the experimental cross-sections.
In leading-order QCD, the momentum fraction of the incident parton can 
be reconstructed from the final-state jet momenta, and therefore a 
direct unfolding of the gluon density is possible \cite{5}. This 
procedure does no longer work beyond leading order, where the 
mass-factorized hard scattering cross-section is a distribution (in the 
mathematical sense) to be convoluted with the parton densities.

For general factorization schemes, the parton-level cross-section 
contains subtractions, thus ``the momentum fraction of the incident 
gluon'' in a naive probabilistic interpretation no longer makes sense.
The physical origin of this phenomenon is initial-state radiation (see 
Fig.~\ref{figref1}c; in this example, the antiquark $\overline{q}$ is 
assumed to be radiated into the forward direction, giving a contribution
to the hard process of Fig.~\ref{figref1}a). The separation of 
the calculable short-distance subprocess from the long-distance physics
of the proton state (whose onset appears in the form of collinear and 
soft divergences of matrix elements) requires the renormalization of 
the parton densities. A change of the factorization scheme amounts to 
a redefinition of both the hard scattering cross-section as well as of 
the parton densities --- in other words, the notion of an ``incident 
parton'' becomes a factorization-scheme dependent concept --- such that, 
to all orders, observable quantities are unaffected. For general 
factorization schemes, the determination of the gluon density has to 
rely on a fitting method.

\section{Two Scenarios at HERA}

\begin{figure}[htbp] \unitlength 1mm
\begin{center}
\begin{picture}(160,79)
\put(0,5){\epsfig{file=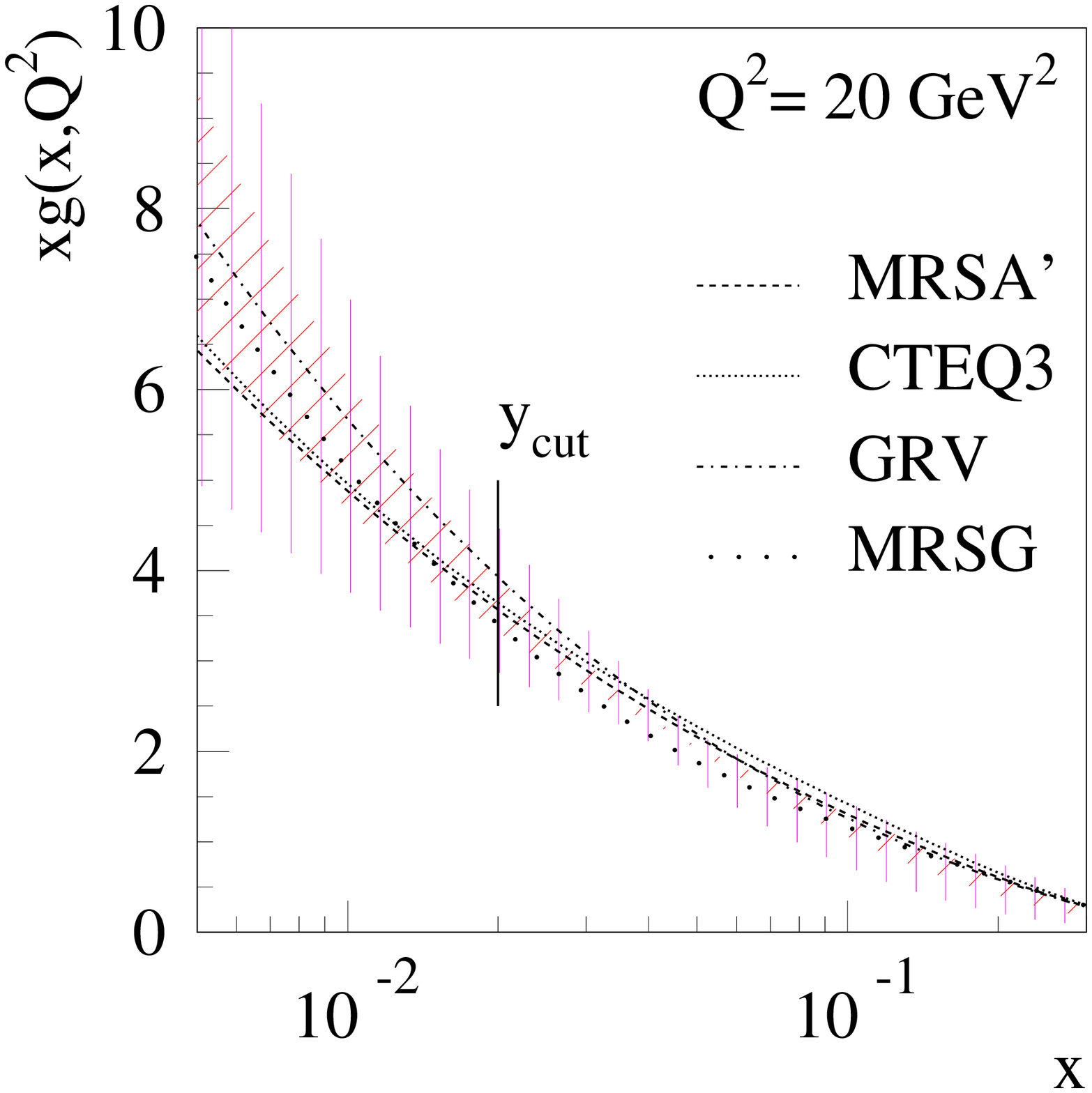,width=8cm}}
\put(80,5){\epsfig{file=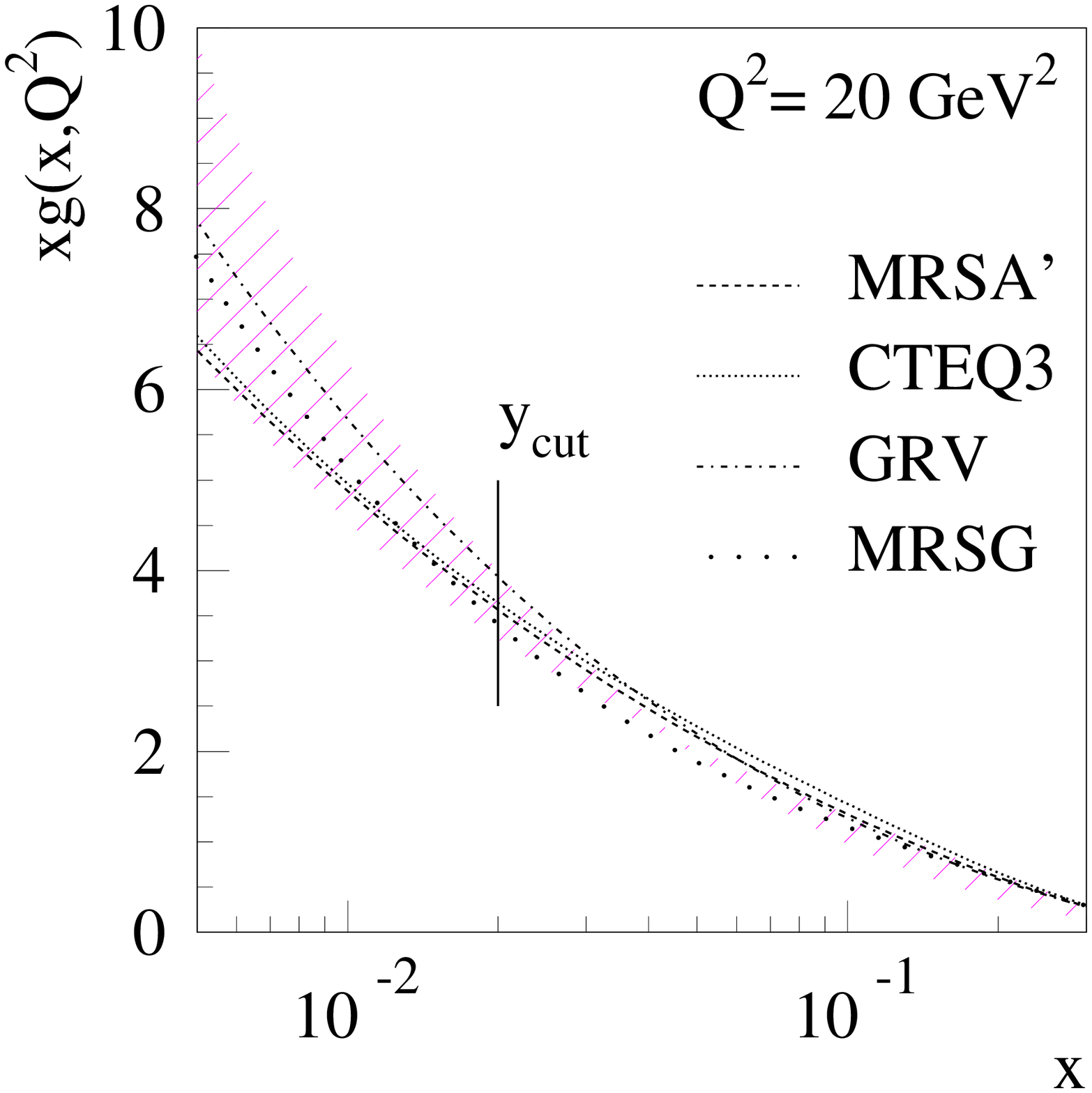,width=8cm}}
\put(22,22){\rm (a)}
\put(102,22){\rm (b)}
\end{picture}
\end{center}
\vspace{-13mm}
\caption[]
{{\it 
Fit result on the proton's gluon density with typical error bands for 
statistical, systematic (and theoretical) errors added in quadrature 
compared with recent parametrizations {\rm \protect\cite{2}}.
(a) for 1994 HERA luminosity, (b) high-luminosity scenario. For details 
see the text.
}}
\label{errorbands}
\end{figure}

In this section, we present the explicit results for a study based on 
jet rates in the modified JADE jet definition scheme \cite{6} with a 
jet cut parameter of $y_{cut}=0.02$. We use the {\tt PROJET} 
next-to-leading-order Monte-Carlo program \cite{7} with the matrix 
elements from Ref.~\cite{8}. The program is applied in a phase 
space region (jets in the very forward direction are excluded) where 
the approximations made in the calculation underlying {\tt PROJET} should be 
justified.
In Fig.~\ref{errorbands}a the error bands for a typical statistical-, 
systematic- and theoretical-error scenario are shown. The employed 
acceptance cuts are discussed in Ref.~\cite{9}. A luminosity of 
about $3\,\mbox{pb}^{-1}$ has been assumed. The inner shaded region 
shows the statistical plus experimental systematic errors, added in 
quadrature. The outer shaded region displays the total error, obtained 
by subsequently adding the theoretical error quadratically,
including a scale variation in the range of $[Q/2,2Q]$.
Only the region with $x>y_{cut}$ is covered by data\footnote{
The use of the modified JADE algorithm restricts the accessible range 
in the gluon momentum fraction to values larger than the jet cut, 
because the invariant mass of the outgoing hadronic system is 
constrained. It is possible to extend this range by using other jet 
algorithms, such as the cone or the $k_T$ schemes.
}, but the extrapolation to smaller~$x$ exhibits a reasonable behaviour 
of the gluon density parametrization.

In Fig.~\ref{errorbands}b it is assumed that, owing to a much 
higher luminosity (of the order of $250\,\mbox{pb}^{-1}$), the 
systematic error can be halved by much tighter cuts leaving the same 
statistical error in the sample, and that the progress in the 
understanding of theoretical uncertainty will allow for neglecting its 
influence against the remaining experimental error. Under these
assumptions the error is reduced dramatically, enabling a discrimination
between different present-day parton parametrizations \cite{2}.
It should be noted that the spread of these parametrizations does 
not represent the real uncertainty on the gluon distribution, due to 
the use of similar data samples and theoretical assumptions.
This emphasizes the discriminating power and constraining effect of 
a DIS jet data sample at HERA with much reduced systematic error in 
the region $0.01<x<0.1$.

\section{Summary and Conclusions}

We have studied the prospects of a direct determination of the 
proton's gluon density via jet rates in deeply inelastic scattering at 
HERA. The application of the Mellin transform method \cite{3} 
allows for an efficient fitting procedure with several parameters, 
the input data of the fit being the jet rates in bins of $Q^2$. 
We have illustrated that high luminosity at HERA and, consequently,  
smaller systematic errors owing to tight acceptance cuts, can permit a 
useful constraint of the gluon density, complementary to (and finally
includible in) global fits. Forthcoming high-statistics runs of HERA 
will also allow for a binning in other variables more closely related 
to the (unobservable) momentum fraction of the gluon, which should have 
a direct impact on the quality of the fit.

\medskip
\medskip
\medskip
\noindent
{\Large \bf Acknowledgements}

\medskip
\noindent
This work was supported in part by the German Federal Ministry for 
Research and Technology (BMBF) under contract No.\ 05 7WZ91P (0).
M.~Hampel gratefully acknowledges support by the Studienstiftung
des deutschen Volkes.


\newcommand{\scs}{\rm}
\newcommand{\bibitema}[1]{\bibitem[#1]{#1}}
\newcommand{\bibbeginshort}{\begin{thebibliography}{XX}}
\newcommand{\bibbeginlong}{\begin{thebibliography}{XXX\,1988\,\,}}

\bibbeginshort 

\bibitema{1}
   W.~Vogelsang and A.~Vogt, 
   Nucl.\ Phys.\ \underline{B453} (1995) 334.

\bibitema{2}
      A.~Martin, R.~Roberts and W.J.~Stirling,
      Phys.\ Lett.\ \underline{B354} (1995) 155; \\
      M.~Gl\"{u}ck, E.~Reya and A.~Vogt,
      Z.~Phys.\ \underline{C67} (1995) 433; \\
      H.L. Lai et al., CTEQ Collaboration,
      Phys.\ Rev.\ \underline{D51} (1995) 4763.

\bibitema{3}
      D.~Graudenz, M.~Hampel, A.~Vogt and Ch.~Berger,
      Z.~Phys.\ \underline{C70} (1996) 77.

\bibitema{4}
      S.~Catani, Y.L.~Dokshitzer and B.R.~Webber,
      Phys.\ Lett.\ \underline{B285} (1992) 291;\\
      B.R.~Webber, J.\ Phys.\ \underline{G19} (1993) 1567.

\bibitema{5} 
      H1 Collaboration, S.~Aid et al., 
      Nucl.\ Phys.\ \underline{B449} (1995) 3.

\bibitema{6}
      D.~Graudenz and N.~Magnussen,
      Proceedings of the 1991 HERA Workshop,
      eds.\ W.~Buchm\"uller and  G.~Ingelman 
      (DESY, Hamburg, 1991), p.\ 261.

\bibitema{7}
      D.~Graudenz, Comput.\ Phys.\ Commun.\ 
      \underline{92} (1995) 65.

\bibitema{8}
      D.~Graudenz,
      Phys.\ Lett.\ \underline{B256} (1991) 518;
      Phys.\ Rev.\ \underline{D49} (1994) 3291.

\bibitema{9}
      M.~Hampel, Ph.D.\ thesis (1996), in preparation.

\end{thebibliography}

\end{document}